\documentclass[final,5p,times,twocolumn]{elsarticle}

\usepackage{graphicx,bm}
\usepackage{amsmath} 
\usepackage{amssymb}
\usepackage{latexsym} 
\usepackage{xcolor}
\usepackage{float}
\usepackage{graphicx}
\DeclareMathOperator{\sgn}{sgn}

\newcommand{\be}{\begin{equation}}
\newcommand{\ee}{\end{equation}}
\newcommand{\bes}{\begin{equation*}}
\newcommand{\ees}{\end{equation*}}
\newcommand{\bea}{\begin{eqnarray}}
\newcommand{\eea}{\end{eqnarray}}
\newcommand{\bi}{\begin {itemize}}
\newcommand{\ei}{\end {itemize}}
\newcommand{\benm}{\begin{enumerate}}
\newcommand{\eenm}{\end{enumerate}}
\newcommand{\bmn}{\begin{minipage}}
\newcommand{\emn}{\end{minipage}}
\newcommand{\bfig}{\begin{figure}}
\newcommand{\efig}{\end{figure}}
\newcommand{\ig}{\includegraphics}
\newcommand{\lnw}{\linewidth}

\newcommand{\bcls}{\begin{columns}}
\newcommand{\ecls}{\end{columns}}
\newcommand{\bcl}{\begin{column}}
\newcommand{\ecl}{\end{column}}

\newcommand{\dr}{{\dot{R}}}
\newcommand{\ddr}{{\ddot{R}}}

\newcommand{\lla}{\left\langle}
\newcommand{\rra}{\right\rangle}
\newcommand{\lal}{\langle}
\newcommand{\ral}{\rangle}

\newcommand{\mum}{\mu\text{m}}

\def\vep{\varepsilon}

\def\al{\alpha}

\def\om{\omega}

\newcommand{\RN}[1]{%
	\textup{\uppercase\expandafter{\romannumeral#1}}%
}

\graphicspath{ {./} }
 
\journal{Ultrasonics Sonochemistry}

\begin{document} 
\begin{frontmatter}
  \title{Machine learning models for the secondary Bjerknes force between two insonated bubbles}

\author{Haiyan Chen, Yue Zeng} 
\address{School of Material and Energy, Guangdong University of
  Technology, Guangzhou, China, 510006}

%\author{Yue Zeng} 
%\address{School of Material and Energy, Guangdong University of
%  Technology, Guangzhou, China, 510006}

  \author{Yi Li\corref{c1}}
\ead{{yili@sheffield.ac.uk.}}
  \cortext[c1]{To whom correspondence should be addressed.}
\address{School of Mathematics and Statistics, University of Sheffield, Sheffield, UK, S3 7RH}

%\maketitle

\begin{abstract} 

%The interaction between two oscillating bubbles in viscoelastic fluids has not
%  received much attention despite its importance in biomedical
%  and other industrial
%  applications. 

  The secondary Bjerknes force plays a significant role in the 
  evolution of bubble clusters. However, due to the complex dependence of the force on multiple
  parameters, it is highly non-trivial to include the effects of this force in the simulations 
  of bubble clusters. 
  %As a matter of fact, the force has mostly been neglected in the state of the
  %art simulations of bubble clusters, or been treated in the most simplistic ways. 
  In this paper, 
  machine learning is used to develop a data-driven model for the secondary Bjerknes force between
  two insonated bubbles
  as a function of the equilibrium
  radii of the bubbles, the distance between the bubbles, the amplitude and the frequency of the 
  pressure. 
  The force varies over several orders of magnitude, which poses a serious challenge for the usual
  machine learning models. To overcome this difficulty, the 
  magnitudes and the signs of the force are separated and modelled separately. A nonlinear
  regression is
  obtained with a feed-forward network model for the \emph{logarithm} of the magnitude, 
  whereas the sign is modelled by a support-vector machine model. 
  The principle, the practical aspects related to the training and validation of the machine models are
  introduced. The predictions from the models are checked against the
  values computed from the Keller-Miksis equations. The results show that the models are extremely efficient
  while providing 
  accurate estimate of the force. The models
  make it computationally feasible for the future simulations of the bubble clusters to include the effects of
  the secondary Bjerknes force. 
  
\end{abstract}


\end{frontmatter}

\section{Introduction \label{sect:intro}}

The 
secondary Bjerknes force \citep{Leighton94,Brennen95} 
is the interaction between 
two bubbles oscillating in a acoustically driven fluid, and it is induced 
by the pressure perturbation radiated from 
the bubbles. 
The force is thought to be important in the evolution of bubble clusters and has attracted
considerable research in the past decades \citep{Crum75, PelekasisTsamopoulos93a, 
PelekasisTsamopoulos93b, 
DoinikovZavtrak95, 
Mettinetal97,
Barbatetal99, Harkinetal01, Pelekasisetal04, 
Yoshidaetal11, Jiaoetal15, Zhangetal16}, which explores 
the effects of nonlinear correction,
multiple scattering, and the coupling with shape oscillation and translation, as well as the experimental
measurement of the force. The asymmetricity of the force is discussed recently in \citep{Pandey19} 
taking into account
higher order nonlinear coupling between the bubbles, which further highlights the complexity of the force.   

Recent experimental evidences \citep{Fanetal14, Lazarusetal17} 
support the importance of the secondary Bjerknes force in the dynamics of micro-bubble
clusters. 
The
collective behaviors of up to $100$ oscillating bubbles are modelled in \citep{Haghietal19} using the coupled Keller-Miksis
equations \citep{KellerMiksis80}. It is found that the interactions between the bubbles can be both constructive and
destructive, and the bifurcation sequences of a system with more bubbles can be much different from a small
one. The research again demonstrates the importance of the interactions between the bubbles which
are manifested as the secondary Bjerknes force. 
The force has been used to 
manipulate bubbles, e.g., as a mean to control micro-devices, which potentially have important 
applications \citep{Ida09, Lanoyetal15, Ahmedetal15}. 
Given that bubble clusters are commonly observed 
in biomedicine, metallurgical industries, food processing, and other applications
(see, e.g., \citep{Brujan11, Bermudezetal11, Roberts14, EskinEskin15}), the modelling of the
secondary Bjerknes force and hence bubble clusters is a question of significant interests.  

Few simulations of bubble clusters so far have employed sophisticated models for the secondary
Bjerknes force. 
Numerical simulations conducted in \citep{Lazarusetal17}, with a simple 
model for the secondary Bjerknes force, 
qualitatively reproduce the experimental observations on the clustering of bubble clouds. 
Similar simplified models are also used  
in the simulations in \cite{Mettinetal99, Parlitzetal99,
Mettin05}, which qualitatively reproduces the formation of the Lichtenberg pattern \cite{Leighton94}.  
These simulations follow the movements of individual bubbles, thus are based on a Lagrangian approach.
Recently a hybrid Lagrangian-Eulerian method is proposed in \cite{MaedaColonius19} where bubble
oscillation is computed, although the secondary Bjerknes force is not explicitly included. 

The past research has yielded considerable physical insights about the secondary Bjerknes force. 
Unfortunately, due to the complexity of the problem, the insights have yet to be translated 
into accurate and computationally efficient models. 
We observe, however, that the complexity of the problem makes it an excellent example for which 
a data-driven approach can be fruitful. Data-driven methods, especially machine learning,
have made tremendous progresses in recent years, as are exemplified and popularized by the success of AlphaGo
\citep{Silveretal17}. The methods have been successfully applied to many physical
and applied
sciences. There is, however, not yet any report of such applications in bubble
simulations.  
The objective of this paper is to use machine learning to build a novel model for the secondary Bjerknes force  
that is more comprehensive than those previously reported, and more generally, introduce this
useful method into the
investigation and modelling of bubbles oscillations. 

  The paper is organized as follows. The dynamical equations for the bubbles are reviewed in
  Section \ref{sect:eq}, where the dependence of the secondary Bjerknes force on relevant parameters are highlighted. 
  The data set for the force is described in Section \ref{sect:data}. 
  Section \ref{sect:ml_model} introduces the relevant machine learning models to be used to
  build the model for the force. The practical aspects of the training and testing of the models are also presented. 
  Additional checks are performed in Section \ref{sect:tests}
  where the efficiency of the models is also assessed. 
  The conclusions are summarized in 
  Section \ref{sect:conclusions}. 

\section{The governing equations \label{sect:eq}}

Let $D$ be the distance between the two bubbles. 
The radius of bubble $i$ ($i=1,2$) is denoted by $R_i(t)$ and its equilibrium radius is $R_{Ei}$. 
The bubbles are driven by a uniform pressure oscillating harmonically in time:
\be \label{eq:pinf}
p_I(t) = p_0 - p_a \sin(\om t) 
\ee
where $p_0$ is the ambient pressure, $p_a$ is the amplitude of the ultrasonic pressure, and $\omega
\equiv 2\pi f$ and $f$ are the angular and linear frequencies, respectively. By using a 
pressure uniform in space, it
has been assumed that $D$ is small compared with the wave length of the pressure wave or the bubbles
are on the same phase plane of a planar pressure wave. 
The fluid has density $\rho$, speed of sound $c$, surface tension $\sigma$ and  
kinematic viscosity $\nu$. 
 
The radii of the bubbles can be described by the Keller-Miksis model
\citep{KellerMiksis80, Brennen95} with additional pressure coupling terms between the bubbles as
introduced in \citep{Mettinetal97}.
Ignoring the time-delay effect, 
the coupling pressure between bubbles $i$ ($i=1,2$) and $j\equiv 3-i$, denoted as $p_{ij}$, is given 
by \citep{Mettinetal97}
\be \label{eq:coupling}
p_{ij}(t) = \frac{\rho}{D} \frac{d R_j^2 \dot{R}_j}{dt}, 
\ee
which is valid when the radii $R_i$ and $R_j$ are much smaller than $D$. 
With $p_{ij}$ included, the equation for $R_i(t)$ becomes \citep{Mettinetal97}:
\begin{align}
  &   2\rho (1-c^{-1} \dot{R}_i)R_i \ddot{R}_i +  
\rho(3-c^{-1}\dot{R}_i)
\dot{R}_i^2  \notag \\
  = &2(1+ c^{-1}\dot{R}_i)
  (p_{wi} - p_I)
  + 2 c^{-1} R_i (\dot{p}_{wi}- \dot{p}_I) \notag \\
  &- 2\rho D^{-1} (2 R_{j}\dr_{j}^2 + R_{j}^2 \ddr_{j}), \label{eq:km_mettin}
\end{align}
where 
\be \label{eq:pw}
p_{wi} = \left(p_0 + \frac{2\sigma}{R_{Ei}}\right) \left(\frac{R_{Ei}}{R_i}\right)^{3k}- \frac{2
\sigma}{R_i} -
 \frac{4 \rho \nu
\dot{R_i}}{R_i}, 
\ee
is the pressure on the outer wall of bubble $i$ and $k$ is the polytropic exponent for the gas
inside the bubble. We note
that other models for the oscillation of coupled bubbles exist in the literature. 
Obviously the machine learning models to be presented below can be used 
with other
models as well. 

The secondary Bjerknes force is defined as the time-averaged pressure exerting on bubble
$i$ due to the oscillations of bubble $j$ \cite{Crum75, Mettinetal97}.  
Let $F_{ij}$ be the notation for this force, 
simple calculation shows that, for small bubbles, $F_{ij}$ can be written as (see, e.g., \citep{Crum75}):
\be \label{eq:bjerknes}
F_{ij}  
= - \frac{\rho}{D^2} \lla V_i \frac{d R_j^2
\dot{R}_j}{dt} \rra =  \frac{ \rho\langle \dot{V}_i \dot{V}_j\rangle}{4\pi D^2} , 
\ee
where $V_i$ is the volume of bubble $i$. 
The pointed brackets represent time
averaging.
In the above expression we follow the tradition where $F_{ij}$ 
is positive when it is attractive. 
The secondary Bjerknes force
factor $f_{ij}$ \citep{Mettinetal97} is defined as
\be \label{eq:bfctr}
f_{ij} \equiv  D^2 F_{ij} = \rho\frac{ \lal \dot{V}_i \dot{V}_j
\ral}{4\pi} .
\ee
In a bubble cluster, $F_{ij}$ is expected to depend not only
on bubbles $i$ and $j$ but also
the other bubbles. Nevertheless,
when the force was considered in the few bubble cluster 
simulations \cite{Mettinetal99, Parlitzetal99, Lazarusetal17} reported so far, 
$F_{ij}$ had all been calculated from 2-bubble systems, where 
the contributions from other bubbles were neglected. Empirical fitting of $F_{ij}$ 
as a function of $D$ was 
used. The dependence of $F_{ij}$ on other parameters have not been
considered. 

For a 2-bubble system, the only secondary Bjerknes force factor is $f_{12}$. $f_{12}$ depends 
on many parameters of the system, including $R_{Ei}$, $D$, $p_a$,
$\omega$, $\nu$, $\rho$, $c$, $\sigma$, and $k$. In the present investigation, we choose water as the medium,
hence fixing $\nu$ at $0.89\times10^{-6}\text{m}^2/s$, $\rho$ at $997\text{kg}/\text{m}^3$, $c$ at
$1497\text{m}/s$ and $\sigma$ at $0.0721 N/\text{m}$. 
An adiabatic process is assumed so that $k$
is fixed at $1.4$, whereas $p_0$ is assumed to be the atmospheric pressure $p_{\rm atm} =
1.013\times 10^5$Pa. 
The objective of the investigation is to model the dependence of $f_{12}$ (hence $F_{12}$) on the
five parameters: $D$,
$p_a$, $\omega$ (or $f$), $R_{E1}$ and
$R_{E2}$.

\section{The data for $f_{12}$ \label{sect:data}}
The machine learning method is used to discover the complicated dependence of $f_{12}$ on 
the system parameters. The method is data-driven and is based on a large data set for $f_{12}$
obtained over a range of values for the five parameters.
The distance $D$ ranges from
$100\mum$ to $1000\mum$ with an
increment of $100\mum$. The pressure amplitude $p_a$ ranges from $40$kPa to $150$kPa with an
increment of $10$kPa. This
range covers both near harmonic and strongly nonlinear aharmonic oscillations. The
forcing frequency $f$ ranges from 20kHz to 40kHz with an increment of 10kHz. 
Both $R_{E1}$ and $R_{E2}$ start at $1\mum$ and end at $10\mum$ with an increment of $2\mum$. 

When the other parameters are held fixed, a bubble pair with radii $(R_{E1}, R_{E2})$ would have the same $f_{12}$
as a pair with radii $(R_{E2}, R_{E1})$. Therefore the parameter combinations with $R_{E1} < R_{E2}$ are
removed from the data set, which leaves in total 5400 combinations of parameter values. Eq. \ref{eq:km_mettin} is 
numerically integrated for each combination 
to obtain the corresponding $f_{12}$. The ode45 solver in MATLAB is used. In each run,
the simulation is run for ten periods of the forcing pressure to allow the
oscillation becomes stationary. The data from the last two periods are used to calculate
the force factor $f_{12}$ according to Eq. \ref{eq:bfctr}.  
The data for $f_{12}$ obtained this way, and the corresponding parameters, form 
the dataset for the development of the machine learning models. In the terminology of machine
learning, a set of values for the five parameters is called a \emph{predictor}, the corresponding
$f_{12}$ is a \emph{response}. 

\section{The machine learning models 
\label{sect:ml_model}}

The secondary Bjerknes force factor $f_{12}$ depends sensitively on the flow parameters. As a
result, the magnitude for $f_{12}$ varies over many orders of magnitude, 
which poses a significant difficulty for the development of the
machine learning models. 

To overcome the difficulty, the data set for
$f_{12}$ is split into two. 
The first one contains $\log_{10} |f_{12}|$, whereas the second one contains the sign of $f_{12}$
($\sgn f_{12}$).
Two machine learning models are built for the two sets separately, and the prediction for $f_{12}$ is
reconstructed from the two models. The first model, given the nature of the data, is a regression
model, which is implemented with a feed-forward neural network (FFNN). 
The data in the second data set are binary (they are either $1$ or $-1$). A 
classification model, the support-vector machine (SVM), is thus used. 
If the predictions from the first and the second models are $y_1$ and $y_2$, respectively,
the prediction for $f_{12}$ is then given by $ y_2 10^{y_1}$.  

Working with $\log_{10}|f_{12}|$ proves to be crucial. Taking the logarithm 
reduces the range of the data, and
as a result, a FFNN can be found to model $|f_{12}|$ (after exponentiation) and hence $f_{12}$ with
good accuracy. 
Without separating the magnitude and the sign and taking the logarithm of the magnitude, 
we failed to find a satisfactory ML model for $f_{12}$. 
The FFNN and the SVM models are now explained. 

\subsection{The feed-forward neural network for $\log_{10} |f_{12}|$\label{sect:ffnn}}

\begin{figure*}[ht]
\ig[width=\lnw]{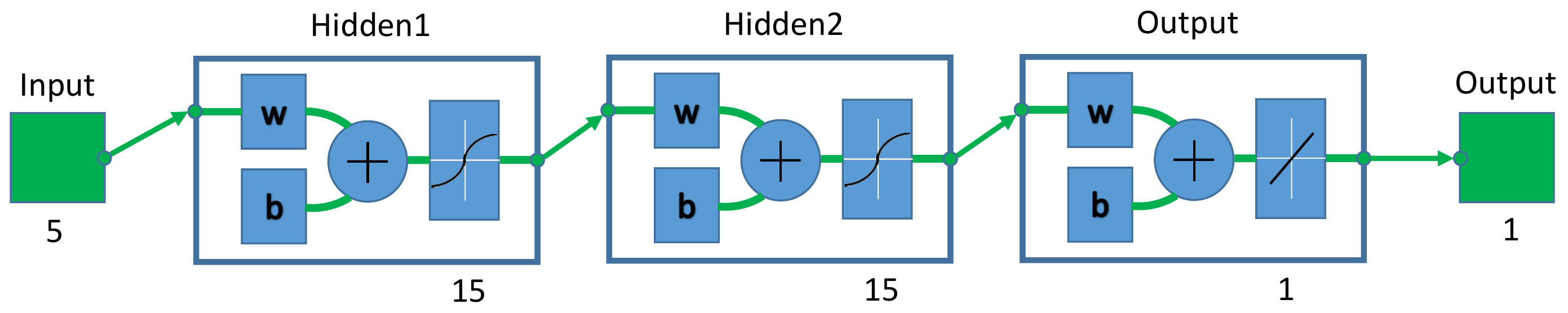}
  \caption{\label{fig:arch} The architecture of the FFNN with two hidden layers having 15 
  neurons on each.}
\end{figure*}

An FFNN \citep{Haganetal14} typically includes an input layer, an out put layer, and several
hidden layers of neurons. The neurons are connected to and receive inputs from those in previous
layers, and similarly, connected to and send outputs to those in later layers. 
Each neuron is defined by an activation function (also known as transfer function), which 
processes the inputs and generates an output. The inputs are combined with suitable weights $w$ and
biases $b$ in the activation function. Fig. \ref{fig:arch} provides a schematic illustration of the
structure of an FFNN. All the components in a blue box forms a neuron. Only one neuron is shown
explicitly in
each layer in Fig. \ref{fig:arch} but in reality there could be many. 
In an FFNN, the number of hidden layers, the number of neurons in each layer, and the 
activation functions are chosen \textit{a priori} and, in practice, mostly
empirically. 
The weights $w$ and biases $b$ are
determined by `training' the network to provide the optimal description of the data in the so-called
`training' dataset. The optimal weights and biases are usually obtained by applying  
optimization algorithms which adjust the weights and biases
iteratively through a process called
back-propagation. For more details on FFNNs and machine learning in general, see, e.g.,
\cite{Haganetal14, Hastieetal09, Goodfellowetal17}. 

\subsubsection{The architecture and the hyperparameters}
MATLAB is used to define, train, validate and test the
network \cite{Ciaburro17}. 
The numbers of layers and neurons and the activation functions are the hyperparameters that should
be decided at the outset. 

For the activation functions, the default setting is adopted, where 
the hyperbolic tangent sigmoid function is used for the neurons in the hidden layers and the linear function is
used in the output layer. 
Even though in the machine learning community rectified linear units (ReLUs) are now recommended for
large scale problems,
a few tests using the ReLUs for the
hidden layers do not show appreciable differences.  

As for the numbers of the layers and neurons, it is known that perfect
regression can be obtained for any dataset if there is no limit to the number of available neurons. 
However, the training may become too expensive and the model too inefficient if too many neurons are
used. Therefore it is desirable to use as few neurons as possible.  
Empirical evidences 
show that, in some cases, the model performance can be improved by 
using more hidden layers \cite{Goodfellowetal17}.
However, there is not yet theoretical justification or guidelines for the optimal choices. 
As a result, we have adopted the following trial-and-error strategy to decide these two hyperparameters, which is
explained briefly here while the numerical evidences is presented later. 
We start
with an FFNN with only one hidden layer, and train it with increasing number of neurons, until
satisfactory performance is obtained. After a number of tests, it is found that consistently
good results can be obtained with $30$ neurons. 
The total number
of neurons is thus fixed at $30$, and networks with different numbers of hidden layers are tested to explore
how the performance can be further improved. 
As demonstrated below
with numerical results, 
%a few FFNNs with two to
%four hidden layers indeed yield better performance than the single layer model.
the best performance is found with two hidden layers and 15 neurons on each. 
This architecture is thus chosen, which is illustrated in
Fig. \ref{fig:arch}. More details are given in Section \ref{sect:NNResults}.

\subsubsection{The training of the FFNN}

With the architecture chosen, 
the weights and the biases in the neurons are then initialized randomly, but they 
have to be adjusted to improve the performance of the
model. This process is called training, in which the dataset for $\log_{10} |f_{12}|$ mentioned in
Section \ref{sect:data} 
is used. For each predictor (i.e., a parameter combination), the FFNN is used to make a prediction for
$\log_{10}|f_{12}|$, which is the response in this model. The prediction  
is compared with the true value obtained as explained in Section
\ref{sect:data}.  
The mean squared error (MSE) between the true and predicted values is used as the performance measure for the
model. 
The
Levenberg-Marquardt algorithm \citep{Haganetal14} is used to optimize the weights and biases
iteratively. In the machine learning terminology, an iteration which scans through all training
data is called an epoch. 

The training is stopped when a certain stopping condition is satisfied. 
One of these conditions is that the magnitude of the gradient of the MSE should be sufficiently
small. However, in our tests, the training is always terminated due to detection of overfitting. 
Overfitting is a phenomenon where the model performs well in training, but makes poor predictions for data
outside of the training dataset. It is a common problem to be avoided when training a neural network. 
In this investigation, cross-validation is used to tackle this problem \cite{Hastieetal09}. Specifically, 
$70\%$ of the dataset for $\log_{10} |f_{12}|$ is randomly chosen to form the training set, 
while $15\%$ is used for validation, and the rest for testing.  
In each epoch, the current FFNN is used to make predictions on the validation data
set. The MSE of the predictions over the validation set is monitored. If the MSE increases 
for $N_o=6$ consecutive
epochs (while the MSE for the training set is still decreasing), then overfitting is deemed to have happened
and the training is stopped. The value $6$ is empirical. If a different $N_o$ is used, one might
need to adjust other parameters to obtain a model with similar
performance. 

\subsubsection{Numerical results \label{sect:NNResults}}
\begin{table}
  \begin{center}
    \def~{\hphantom{0}}
    \begin{tabular}{cl}
      \hline
      \hline
      Architecture & $N_n$ in each layer\\
      \hline
      1 & 30 \\
      2 & 20, 10 \\
      3 & 10, 20 \\
      4 & 15, 15 \\
      5 & 5,~ 10, 15 \\
      6 & 10, 10, 10 \\
      7 & 10, 10, 5,~ 5 \\
      8 & 10, 5,~ 5,~ 5,~ 5 \\
      \hline
      \hline
    \end{tabular}
    \caption{The number of neurons ($N_n$) in each layer for each network architecture. \label{tab:arch}}
  \end{center}
\end{table}

\bfig[ht]
\centering
\ig[width=\lnw]{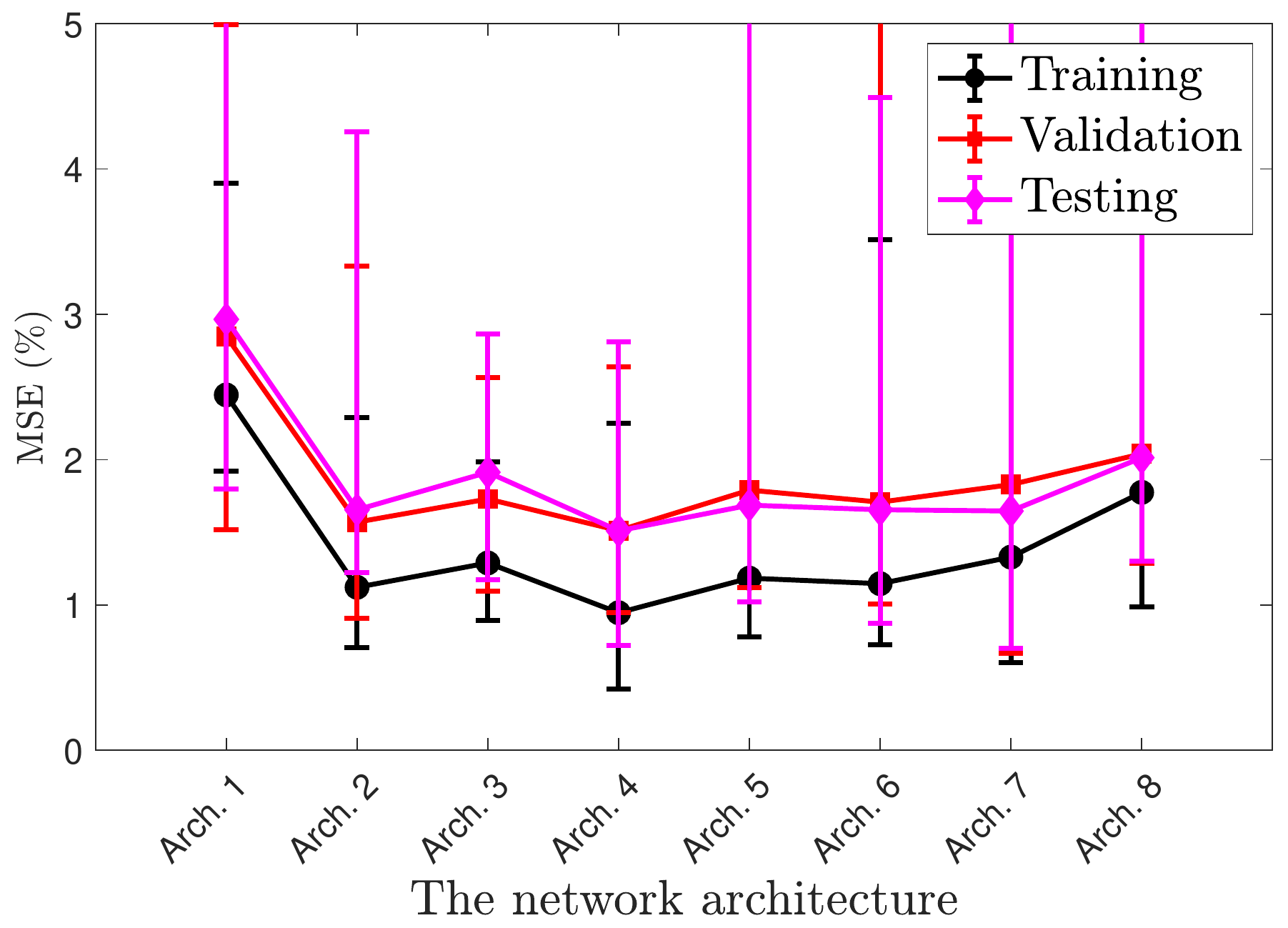}
\caption{\label{fig:NNmse} The median and range of the optimal MSE obtained with different FFNN
models in 32 runs.}
\efig
\bfig[ht]
\centering
\ig[width=\lnw]{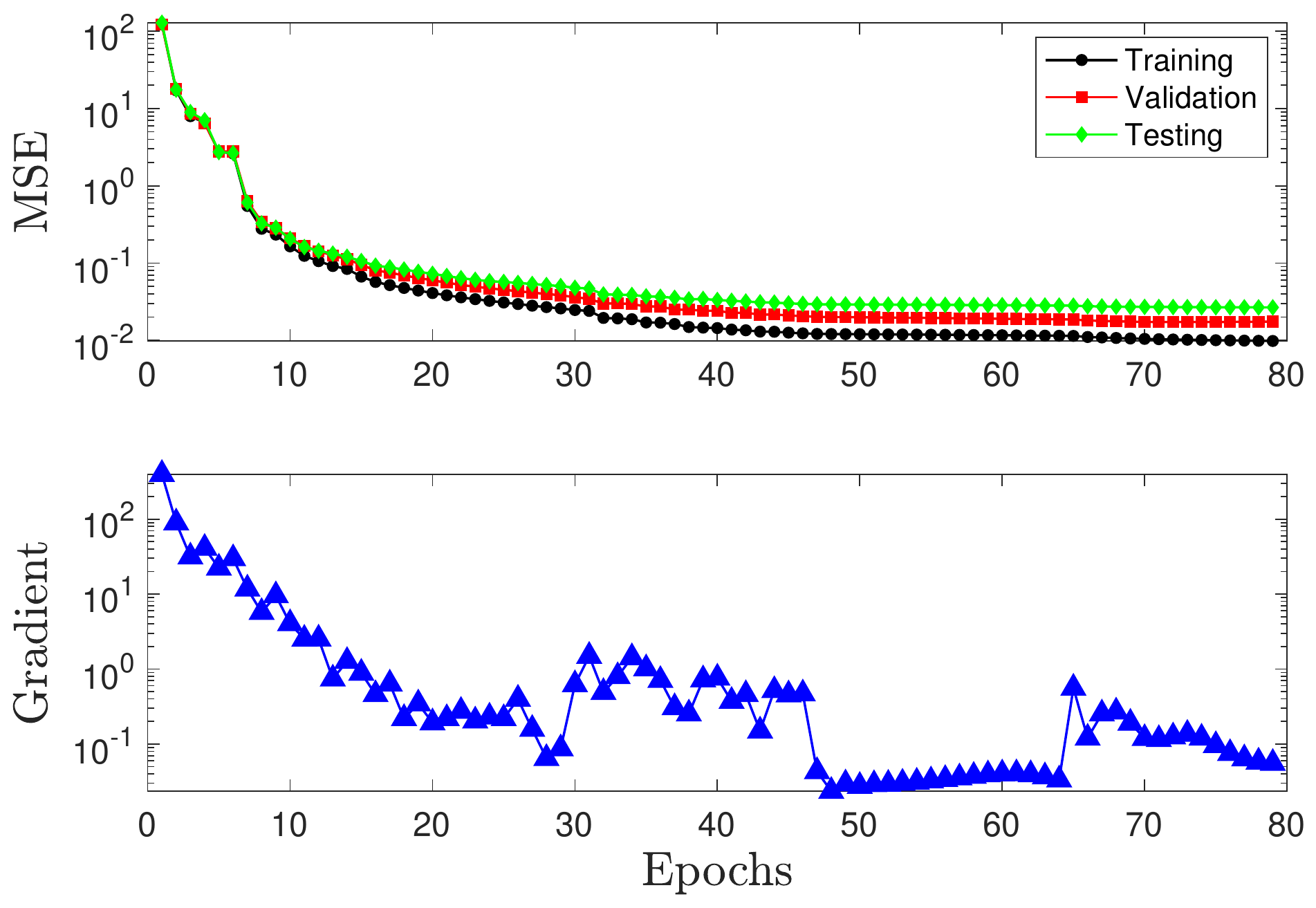}
\caption{\label{fig:NNperf} The changes of the MSE and the gradient with the epochs for a model with
the fourth architecture.}
\efig

The architectures of the FFNN models tested in this investigation are summarized in Table \ref{tab:arch}. For
each architecture, the training is run $32$ times with random initialization. $32$ models are thus
produced, which
are slightly different from each other even if they have the same architecture. 
The 32 optimal validation MSEs for these models obtained
from the training are calculated. The median values are plotted with the symbols in Fig. \ref{fig:NNmse}
for all 8 architectures, as well as the maximum and minimum values, which are given by the error bars.
The result for architecture 1 
shows that, with 30 neurons, the validation MSE can be kept under $5\%$ in all runs. This
observation has
been the basis to choose 30 as the total number of the neurons. 
Fig.
\ref{fig:NNmse} shows clearly that the MSE can be reduced by using multiple hidden layers.
Architecture 4, which has two hidden layers with 15 neurons in each, displays the best performance.
Further increasing the number of hidden layers appears to somewhat degrade the performance. Based on
these results, architecture 4 has been chosen to obtain all the results to be presented below. 
As an illustration, Fig. \ref{fig:NNperf} shows how
the performance and the gradient of the model improves by the training. 

%The eight architectures have been chosen heuristically, based on the following intuitions. Firstly, the number of
%neurons in a layer should not be too small compared with those in other layers, because, otherwise,
%this layer could become the bottle neck that limits the performance. 

\bfig[ht]
\centering
\ig[width=0.9\lnw]{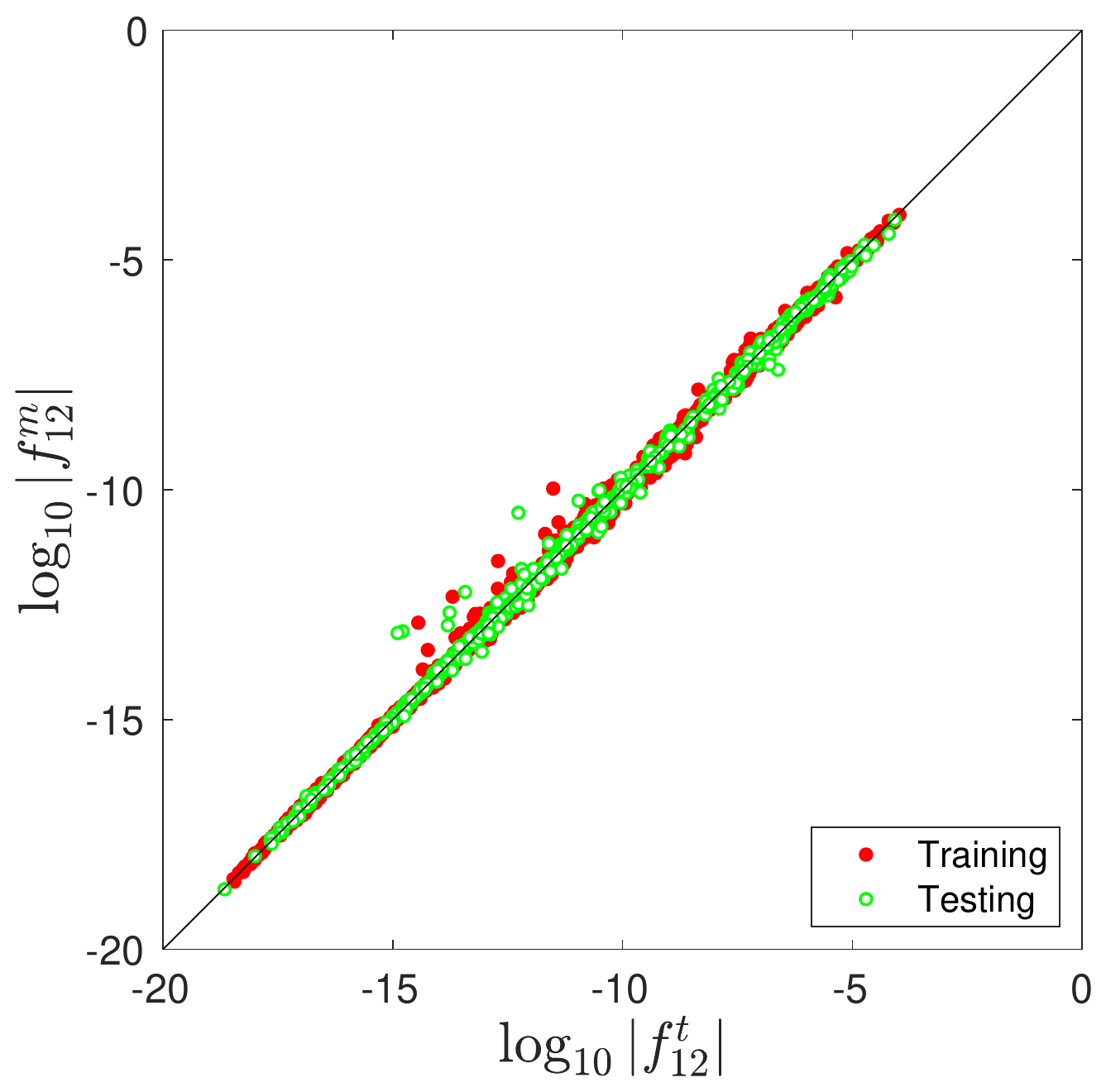}
\caption{\label{fig:NNreg} The regression of the training (filled circles) and the testing (empty
circles) data.}
\efig

\bfig[ht]
\centering
\ig[width=\lnw]{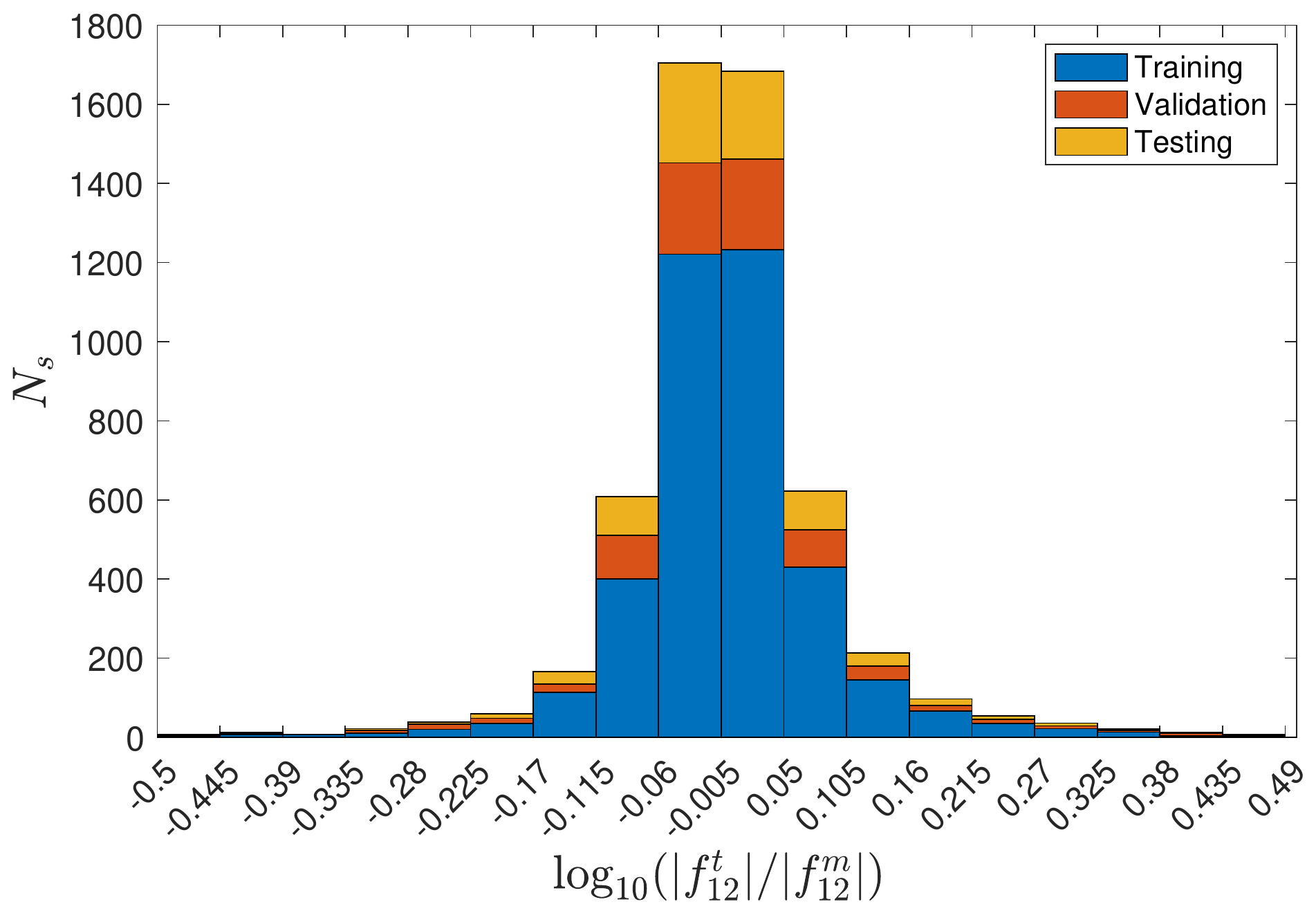}
\caption{\label{fig:NNerr} The bar chart for the error $\vep_a \equiv \log_{10}|f_{12}^t| - \log_{10}|f_{12}^m|
\equiv\log_{10}(|f_{12}^t/f_{12}^m|)$ for the training (bottom bars), validation (middle bars) and
testing (top bars) data. The height of
a bar represents
the number of samples $N_s$ in the bin.}
\efig

\bfig[ht]
\centering
\ig[width=\lnw]{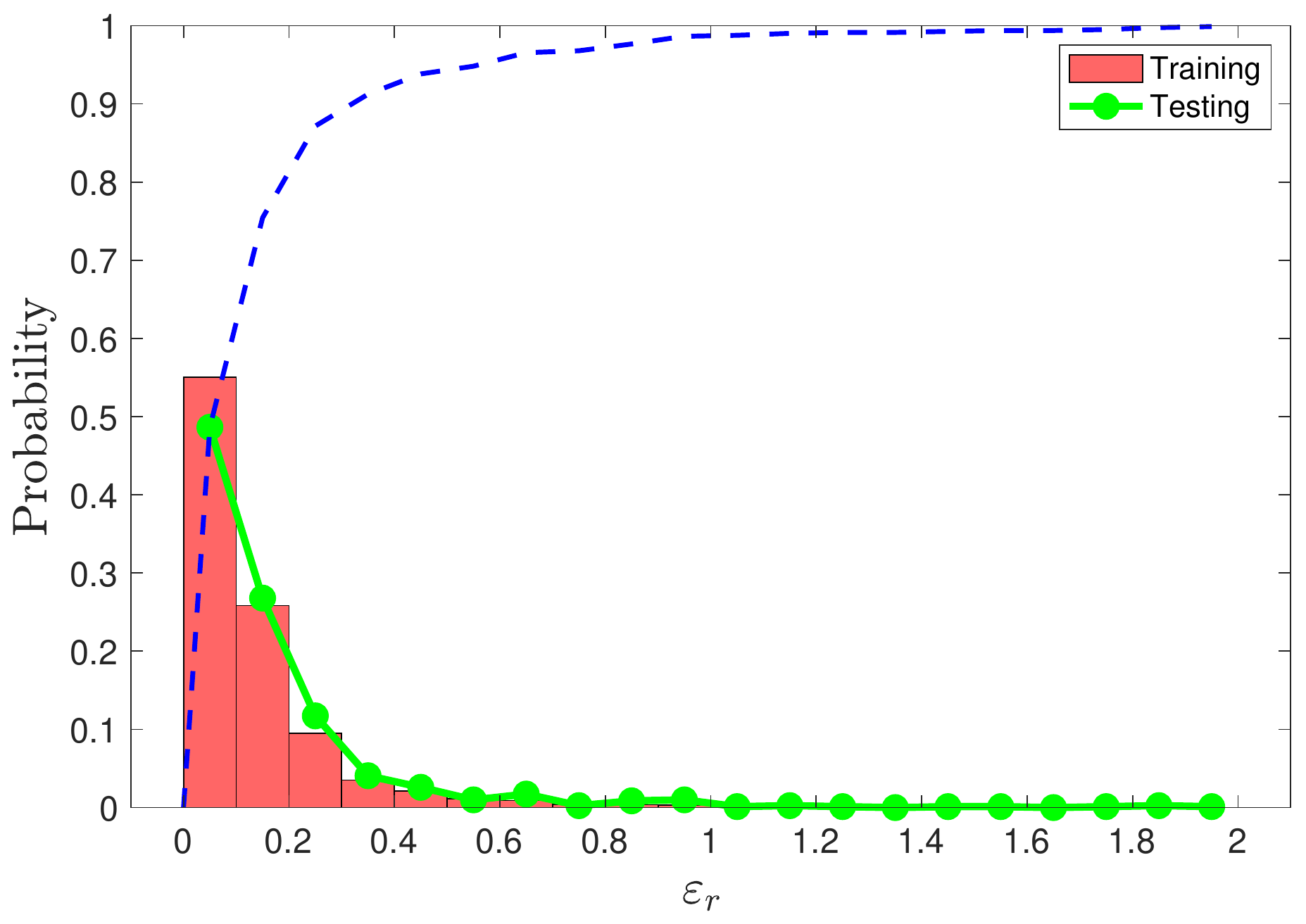}
\caption{\label{fig:NNf12err} The probability distribution for the relative error $\vep_r \equiv |(|f_{12}^t| -
|f_{12}^m|)|/|f_{12}^m|$ for the training (bars) and the testing (filled circles) sets. 
The $Y$-axis is the probability for
$\vep_r$ in each bin. The blue line is the cumulative probability for $\vep_r$ for the testing data.}
\efig

The next a few results provide fuller description of the errors. 
Fig. \ref{fig:NNreg} shows the regression of the training and the testing data, where superscripts $t$ and $m$ 
are used to denote the true and modelled values,
respectively. Excellent regression is obtained for both datasets. 
The coefficient of determination $R$ is more than $99\%$ in both cases. 

The bar chart in Fig. \ref{fig:NNerr} shows the distribution of the absolute error for
$\log_{10}|f_{12}|$ defined by 
$\vep_a \equiv \log_{10} |f^t_{12}| - \log_{10}|f^m_{12}|$.
Given that the values for $\log_{10}|f_{12}|$ approximately range between $-20$ and $-5$ (c.f. Fig. \ref{fig:NNreg}), 
the error on a
large majority data points is very small. 

The error in $\log_{10}|f_{12}|$ is small enough that the
magnitude $|f_{12}|$ itself is also accurately modelled. 
Plotted in Fig. \ref{fig:NNf12err} is the relative
error for $|f_{12}|$, which is defined as $\vep_r\equiv 
||f_{12}^t| -
|f_{12}^m||/|f_{12}^m|
$ and is related to $\vep_a$ by $\vep_r = |10^{\vep_a} - 1|$. 
There are a small
number of
extraneous samples where $\vep_r$ can be more than $100\%$, 
but, as expected, the majority
of the samples have small errors.   
The cumulative probability
distribution, plotted with the dashed line, shows that more than $90\%$ samples in the testing set have relative
errors smaller than $30\%$, and about $85\%$ smaller than $20\%$.  
Comparing the errors on the training set and the testing set, 
it is only slightly more probable to observe larger errors on the testing data, which demonstrates
that the model generalizes well.

The above results show that the FFNN model (with $15$ neurons on each of the two hidden layers) can provide 
an accurate model for not only $\log_{10}|f_{12}|$ but also $|f_{12}|$. In terms of the training
cost, 
typically less than $100$ epochs are needed to achieve
the accuracy depicted in Figs. \ref{fig:NNreg}-\ref{fig:NNf12err}, which takes less than one
minute to compute on a modern laptop. 

\subsection{The support-vector machine model for $\sgn f_{12}$ \label{sect:svm}}

In its most simple form, the SVM \cite{Hastieetal09} is an algorithm that finds the optimal line
that separates two clusters of
points on a plane, hence classifying the points into two classes. In order to introduce some
necessary basic concepts, 
its formulation is briefly explained. 
It is assumed that a set of $N$ points $x^j =(x^j_1, x^j_2)\in R^2$ $(j=1,2,...,N)$ is given  as the training set. 
The points are divided into two groups, the positive and negative classes. For simplicity, 
the data are assumed to be separable, i.e., it is assumed 
that a straight line can be found to separate the two
classes, and that the classifications are known and labelled by $1$ and $-1$, respectively. 
The classification of point $x^j$ is recorded by $y^j\in\{-1,1\}$.

The optimal separating line is defined as the line that
separates the two groups whilst leaving the largest margins on both sides of the line. 
Let the equation for the line be $f(x) \equiv w^T x + b = 0$, 
where $w\in R^2$ and $b\in R$. 
It can be shown that the best separating 
line is the solution of the following optimization problem \cite{Hastieetal09}: find $w$ and $b$ that minimize the
objective $||w||^2$ 
such that for all data points $(x^j,y^j)$,
\be \label{eq:svmconstr1}
y^j f(x^j)\ge 1. 
\ee
The optimal objective maximizes the margins on the two sides of the line. The
constraints ensure that the points are found on the correct sides of the separating line. 

%(TODO: is this paragraph necessary?)
%Usually the margin on the two sides of the line is non-zero. A straight strip with the separating 
%line at the center can be drawn up.
%The support vectors are the points $x^j$ on the boundaries of the strip, for which $y^j
%f(x^j)=1$, hence the name of the method. (TODO: double check) 

The optimal solution is used to classify new data (not in the training set) 
in the following way. Let $\hat{w}$ and $\hat{b}$ be the 
optimal solution, and $\hat{f}(x) \equiv \hat{w}^T x +
\hat{b}$. 
The classification
of a data point $x$ is given by its label $\sgn \hat{f}(x)$. 

%TODO: is this part necessary?
%The classification score for $x$ is defined as $\hat{f}(x)$, which
%is the signed distance from $x$ to the separating line. 
%A positive score for a class indicates that $x$ is predicted (rightly or wrongly) to be in that class. 
%A negative score indicates otherwise.

In the more general cases where the data are not separable, two modifications to the above method
have been introduced. Firstly, one may allow a small number of points to be
mis-classified, a practice termed using soft margin. 
Mathematically, this method amounts to replacing Eq. \ref{eq:svmconstr1} by 
\be \label{eq:svmconstr2}
y^j
f(x^j) \ge 1 - \xi^j 
\ee
for $\xi^j\ge0$. Meanwhile, a penalty term is added to the objective function to limit
the magnitude of $\xi^j$. For this investigation, the term takes the 
form of $C\sum_{j=1}^N \xi^j$ with $C$ being a parameter
called the box constraint. 
A larger $C$ introduces larger penalty,
hence reduces $\xi^j$'s magnitudes, which means fewer mis-classifications are allowed. No
mis-classification is allowed when $C\to \infty$. 

Secondly, one may use a nonlinear curve to separate the data. The idea is implemented by using a
function $f(x) = w^T h(x) + b$ as the separating curve, where $h(x)$ is a non-linear function that
transforms the separating line to a nonlinear separating curve. 

The SVM model obviously is also applicable for higher dimensional problems. 
It is used in this investigation, but the solution is based on
its dual formulation, because the dual problem is convex and guaranteed to converge to the global
minimum \cite{Cristianinietal14}. 
Letting $\alpha^j$ be the dual variable corresponding to the constraint given in Eq.
\ref{eq:svmconstr2}, the dual problem can be written as 
\be \label{eq:dualcost}
\min ~~ \frac{1}{2} \sum_{j=1}^N \sum_{k=1}^N \al^j \al^k y^j y^k G(x^j, x^k) - \sum_{j=1}^N \al^j 
\ee
subject to the constraints
\be \label{eq:dualconsts}
\sum_{j=1}^N \al^j y^j = 0, \quad 0\le \al^j \le C. 
\ee
The bivariate function $G$ in Eq. \ref{eq:dualcost} is the kernel function corresponding to the
nonlinear function $h(x)$. For this investigation, a Gaussian kernel is used where $G(x^j, x^k) =
\exp(-||x^j-x^k||^2/\sigma^2)$ with $\sigma$ being the kernel scale. 
%The dual formulation corresponds to a penalty term $C\sum_{j=1}^N \xi^j$, 
%with the parameter $C$ appearing in Eq. \ref{eq:dualconsts}. $C$ is called the box
%constraint parameter, or simply the box constraint. 
%A larger $C$ introduces larger penalty,
%hence reduces $\xi^j$'s magnitudes, which means fewer mis-classifications are allowed. No
%mis-classification is allowed when $C\to \infty$. 

The dual problem is solved with the sequential minimal optimization
(SMO) algorithm. The SMO is an iterative descent algorithm. The convergence is monitored by the
gradient of the objective function (given in Eq. \ref{eq:dualcost}) with respect to $\al^j$,
specifically, by 
the difference between the gradient components corresponding to the maximal upper and lower violation of
the feasibility conditions
\citep{Cristianinietal14, Fanetal05}. The iteration is deemed converged when this difference is
smaller than a tolerance $\delta$. 
Once the optimal solution for the dual problem is found, $\hat{w}$ and $\hat{b}$ can then be found
from $\al^j$ using the
Karush-Kuhn-Tucker conditions, which are then used to classify a new data point. 
For more details of the SVM methods and the algorithms, see
\citep{Hastieetal09, Cristianinietal14}.

\bfig [ht]
\centering
\ig[width=0.9\lnw]{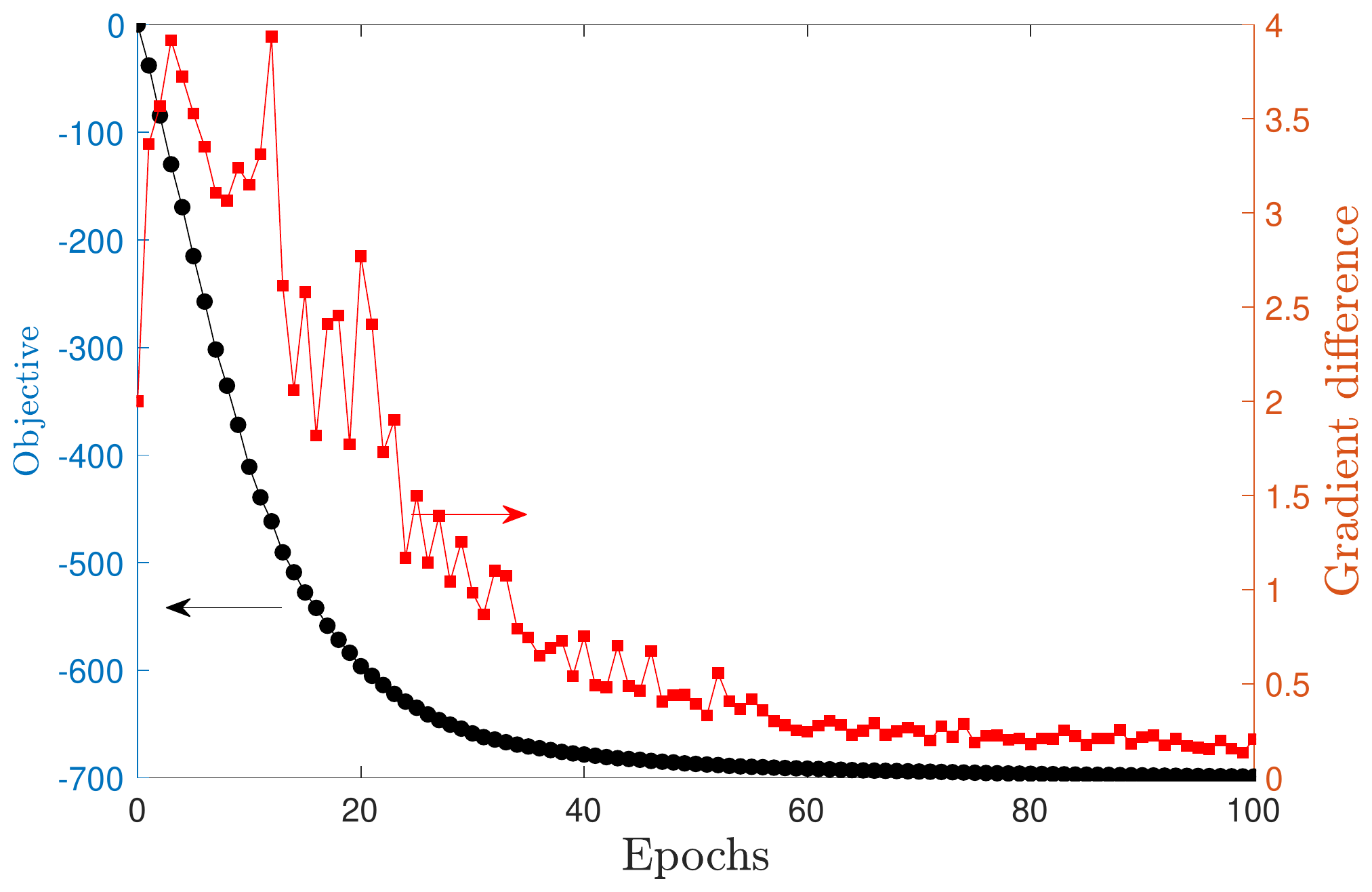}
\caption{\label{fig:svmjcstdgrad} The objective function and the gradient difference as functions of
the training epochs in a typical training session. Only the first 100 epochs are shown.}
\efig

The SVM model is applied to classify the data for $\sgn f_{12}$, using 
the MATLAB command fitcsvm which implements the above algorithms. 
In this case, $x$ is a five dimensional vector, i.e., $x = (D, R_{E1}, R_{E2}, p_a, \om)^T$
and the label $y$ is $\sgn f_{12}$. The number of data points is $N=5400$. The data are
standardized 
when they are fed into the training algorithm. 
Specifically, the mean is removed from the data which are then rescaled by the standard deviation. 
Fig. \ref{fig:svmjcstdgrad} illustrates the decay of the gradient difference as well as
the objective function in Eq.
\ref{eq:dualcost} in a typical training process.  

\bfig [ht]
\centering
\ig[width=0.9\lnw]{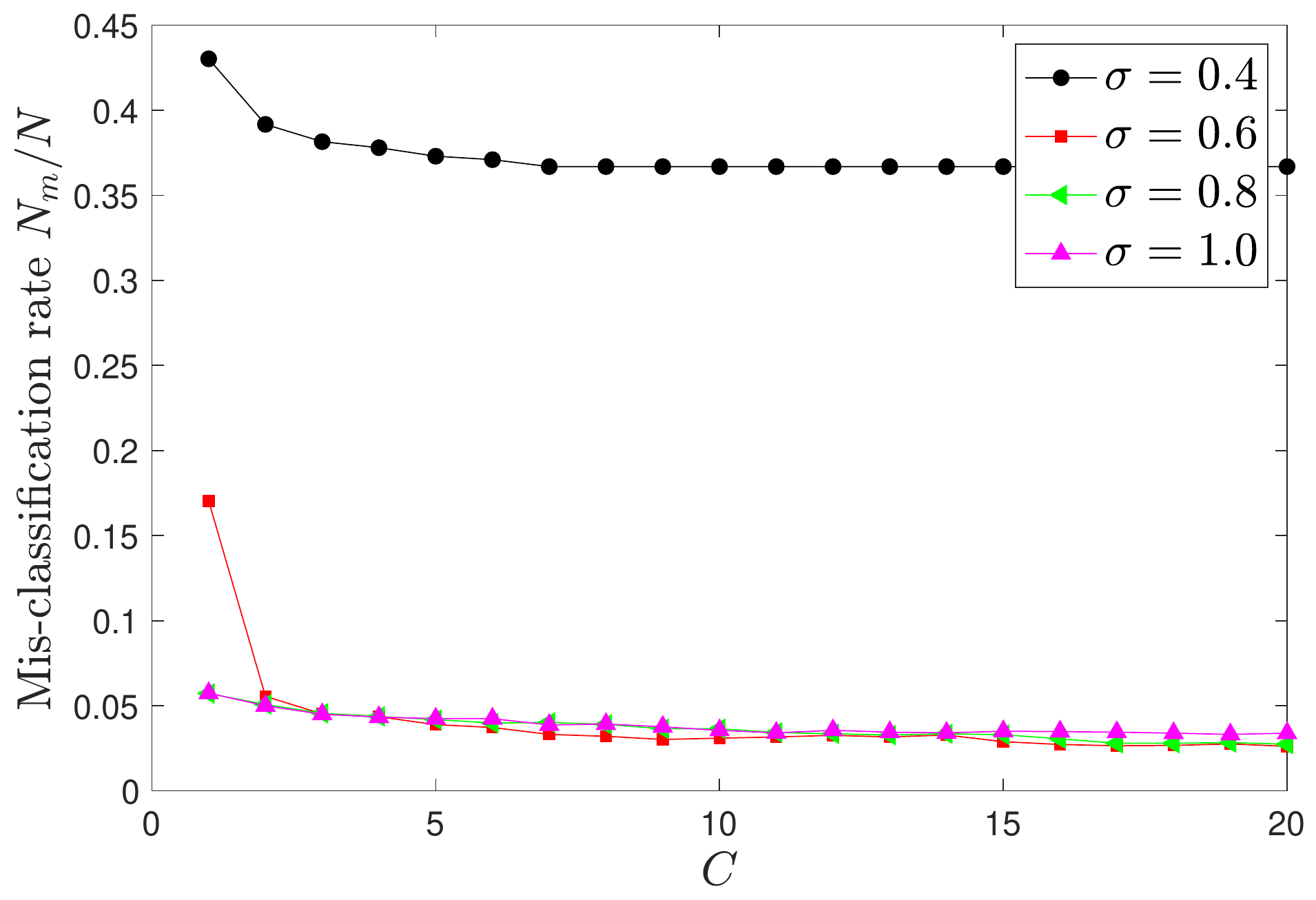}
\caption{\label{fig:svmmiscal} The mis-classification rate for different kernel scales $\sigma$ and box
constraints $C$.}
\efig

The adjustable
parameters in the model are the box constraint $C$, the kernel scale $\sigma$, and the tolerance
$\delta$. 
Tests with $\delta=10^{-3}$ and $10^{-4}$ show essentially the same results. Therefore $\delta =
10^{-4}$ has been used. 
When the separation boundary has a complicated shape, a small $\sigma$ is required to 
model the fine features of the boundary. 
However a too small
$\sigma$ may limit the ability of the model to capture the large scale features in the distribution of the
data. The choice of $C$ depends on the accuracy of the data for $f_{12}$. 
If the data for $f_{12}$ likely contain large errors, it is not meaningful to insist perfect
classification. The physical model being used here (Eq. \ref{eq:km_mettin}) has been derived
with a few simplifying assumptions.  
Therefore, it is appropriate to 
use a large $C$ to limit mis-classification, 
although it is not necessary to achieve zero mis-classification. 

With the above discussion in mind, the mis-classification rate has been calculated for several
different kernel scales $\sigma$ and box constraints $C$. 
A point $x^j$ is mis-classified if $ y^j \hat{f}(x^j)<1$ (c.f. Eq.
\ref{eq:svmconstr2}). The
mis-classification rate is the ratio of the number of mis-classified points, $N_m$, to the total number
$N$. The results are plotted in Fig. \ref{fig:svmmiscal}. The mis-classification rate reaches
an approximate plateau quickly when $C$ increases. The results for $\sigma = 0.4$ are clearly much
worse, whereas small mis-classification is found for other $\sigma$'s when $C$ is sufficiently large.
These observations demonstrate the robustness of the classification scheme as long as $\sigma$ is
not too small. The smallest 
mis-classification of $2.61\%$ is found at $(\sigma, C)=(0.6,20)$, which is considered sufficiently
small. Therefore $C=20$ and $\sigma=0.6$ are
used in what follows. 

\bfig [ht]
\centering
\ig[width=0.9\lnw]{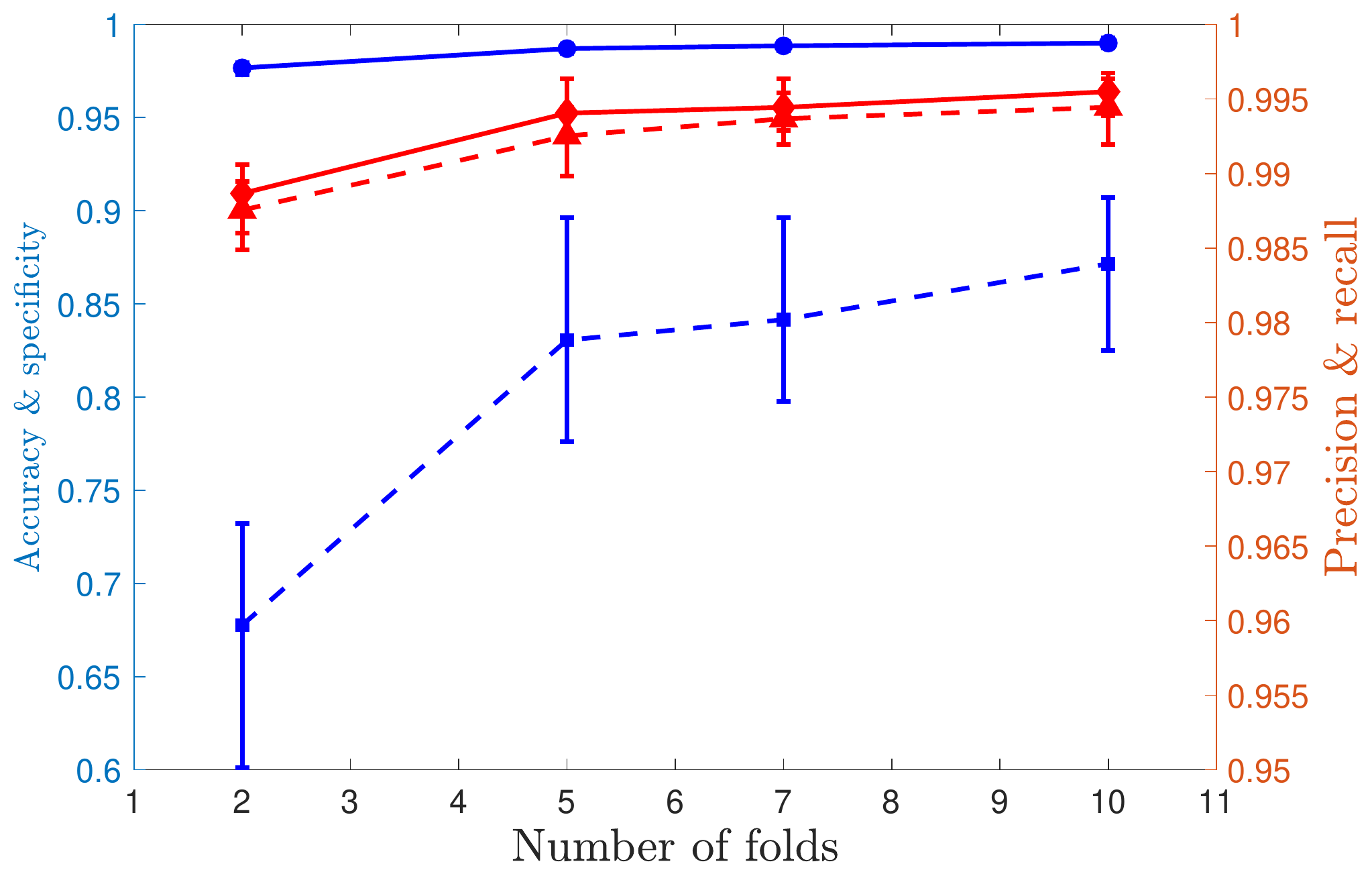}
\caption{\label{fig:svmkfLoss} The left axis: the median accuracy (solid line with circles) and the
median specificity (dashed line with squares).
The right axis: the median precision (solid line with diamonds) and the median recall (dashed line with triangles). 
Found in an ensemble of $32$ runs. 
The error bars show the maximum and minimum.}
\efig

The robustness of the model is assessed in Fig. \ref{fig:svmkfLoss}
using $k$-fold cross-validation \cite{Hastieetal09}. In $k$-fold cross-validation, the dataset is
divided randomly into $k$ equal sets, and $k$ models (called the partitioned models) are trained. 
The $i$th ($i=1,2,...,k$) dataset is called the $i$th test fold, whereas
the other $k-1$ sets form the $i$th training fold. The $i$th partitioned SVM model is trained on the
$i$th training fold and evaluated on the $i$th test fold. The overall assessment is based on 
performance indices averaged over the $k$
partitioned models. 

The most commonly used performance indices are
the accuracy, the precision, the recall and the specificity. For each predictor, the model response could either be
positive or negative; in either case, it could be either true or false.  As a result, the
response falls in one of four
categories: a positive response could be a true positive (TP) or, coming erroneously from a predictor
in the negative
class, a false positive (FN), whereas a
negative response could be true negative (TN) or false negative (FN). Let $N_{TP}$, $N_{TN}$,
$N_{FP}$ and $N_{FN}$ be the numbers of TP, TN, FP, and FN, respectively. The accuracy is defined as 
\be \label{eq:acc}
\frac{N_{TP} + N_{TN}}{N}, 
\ee
which simply gives the percentage of correct predictions in both classes. 
The precision, recall and specificity are, respectively, defined as
\be
\frac{N_{TP}}{N_{TP} + N_{FP}},
\frac{N_{TP}}{N_{TP} + N_{FN}}, \text{ and } \frac{N_{TN}}{N_{TN}+N_{FP}}.
\ee
The precision tells us the probability of a positive response being correct; the recall is 
the probability of the positives being correctly identified as positive, while the specificity gives the probability 
of the negatives being correctly identified as negative \cite{Hastieetal09}.  

Due to the randomness in data partition, the indices averaged over the $k$ partitioned models may fluctuate if
the cross-validation is conducted multiple times. Therefore, we repeat the validation $32$ times to find the
medians and ranges of the indices. 
The results are plotted in Fig. \ref{fig:svmkfLoss}. It is clear that the accuracy, the precision,
and the recall 
of the models are consistently high (more than $98\%$ for all of them), although 
they drop slightly with the number of folds while the ranges increase slightly (the error bars for
the accuracy are too narrow to see on the figure). With fewer
folds, the number of data points in each fold, hence in the training set, is smaller. Thus the performance of the
partitioned models are expected to somewhat deteriorate. The results for accuracy show that more than $98\%$
data points, with either negative or positive $f_{12}$, are classified correctly. 
Meanwhile, the result for the precision shows that
there is a $98\%$ chance that $f_{12}$ is indeed positive when the model predicts
so, and the result for the recall shows that there is a $1-98\% = 2\%$ chance that
$f_{12}$ is not predicted to be positive
when it is actually positive. 

Fig. \ref{fig:svmkfLoss} shows that the median specificity and its 
range display behaviours similar to those of the accuracy, the precision and the recall, 
but there is stronger dependence 
on the number of folds, and
its range is wider and the median is smaller. 
The median recall increases with the fold number and reaches approximately $85\%$, which
implies there is a $1-85\% = 15\%$ chance that
$f_{12}$ is not predicted to be negative
when it is actually negative. 
The specificity is relatively low compared with the other indices, but this observation does
not necessarily reflect poorer performance regarding the samples with negative $f_{12}$. 
Rather, this behaviour is 
due to the fact that there is only about $3.4\%$ data points on which
$f_{12}<0$, thus mis-classification has an outsized impact. 

The results presented in this subsection show that, with kernel scale $\sigma = 0.6$, 
box constraint $C=20$, and tolerance $\delta =
10^{-4}$, the trained SVM classifier can effectively model the
distribution of the signs
of $f_{12}$. 

\section{Efficiency and accuracy of the combined model\label{sect:tests}} 

\bfig[ht]
\centering
\ig[width=0.8\lnw]{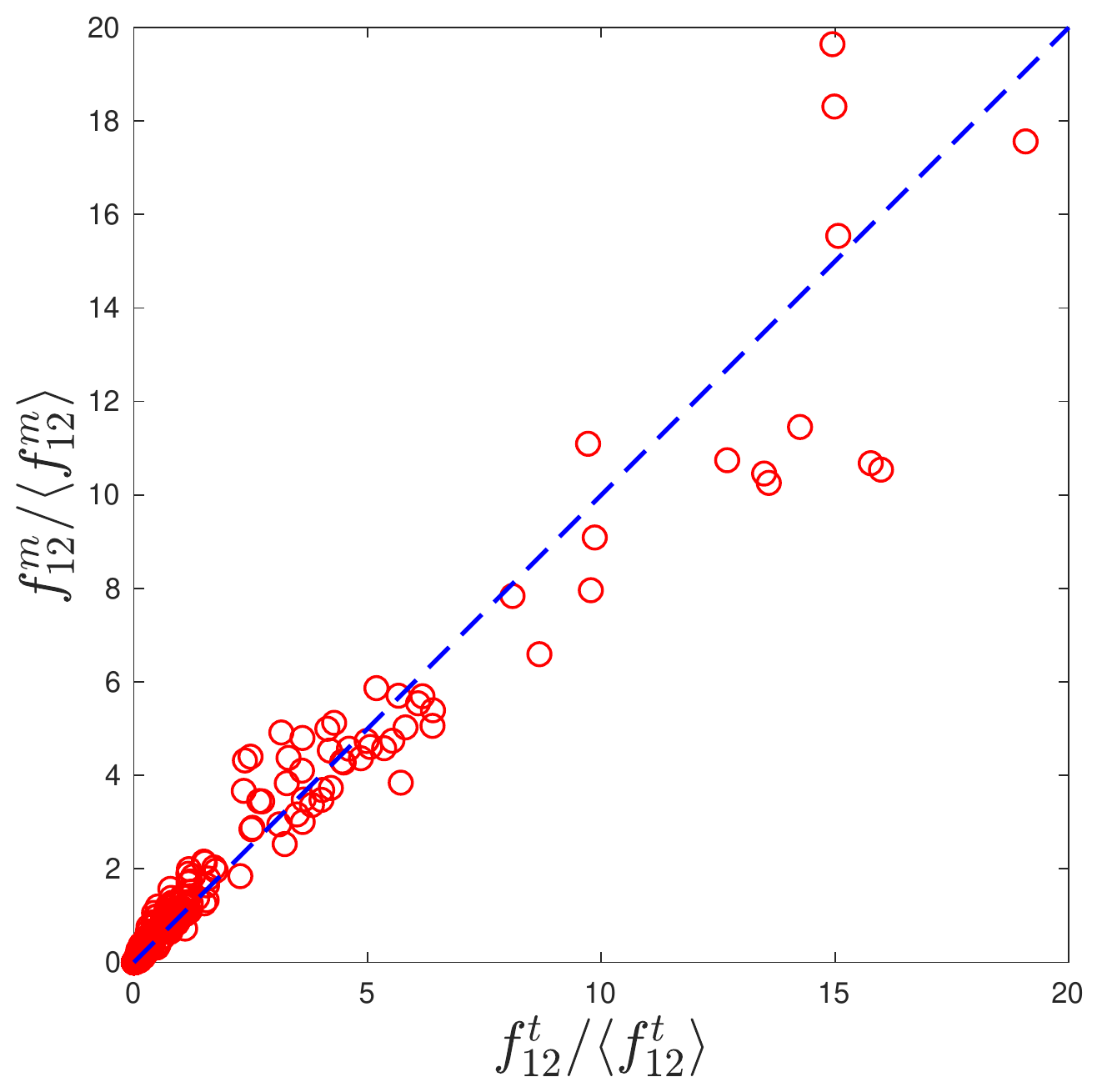}
\caption{\label{fig:cmpscat} The scatter plot for $(f_{12}^t/\langle f_{12}^t\rangle,
f_{12}^m/\langle f_{12}^m\rangle)$, where $\langle \cdot \rangle$ denotes the averaging over the
dataset.  }
\efig
\bfig[ht]
\centering
\ig[width=0.8\lnw]{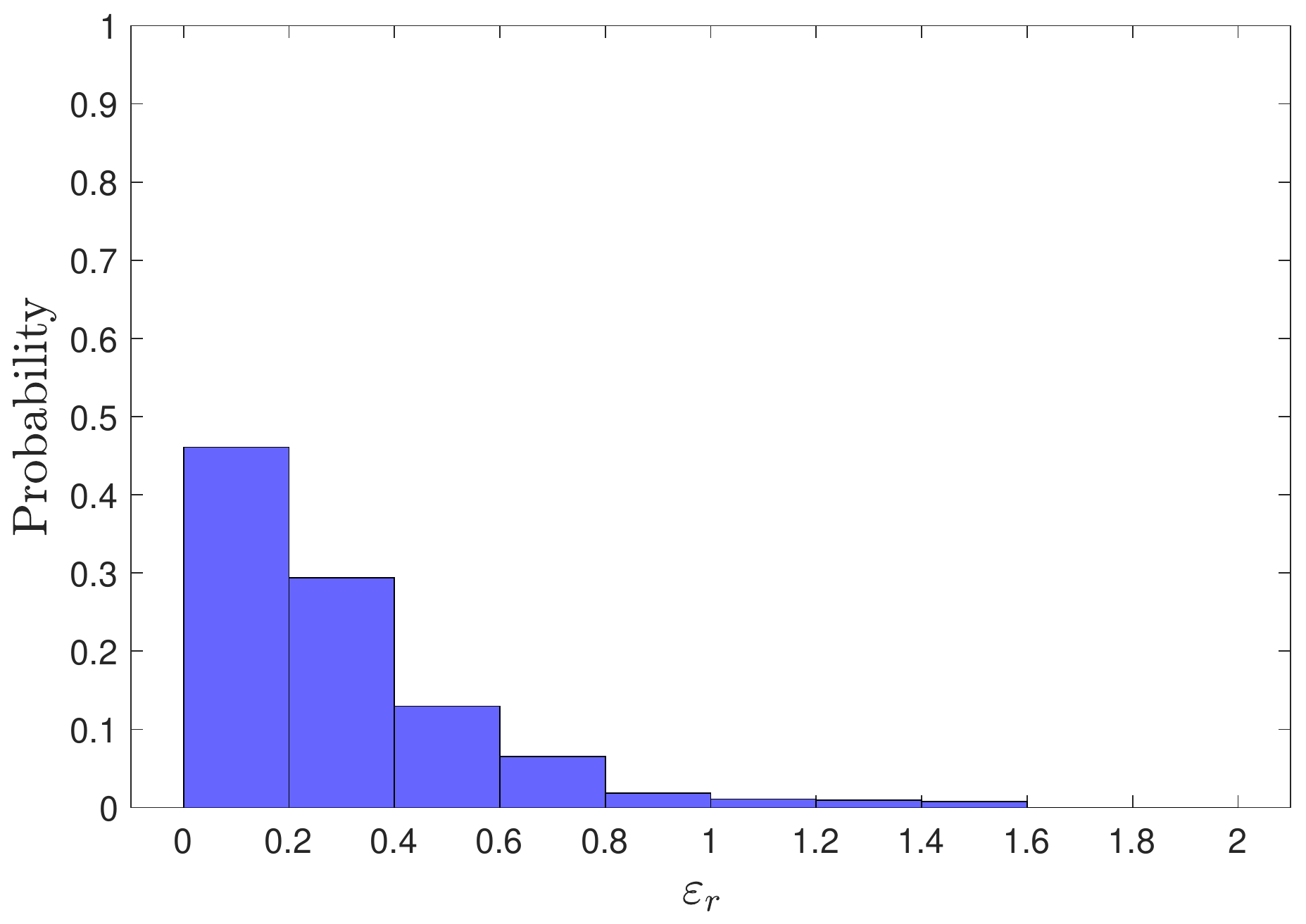}
\caption{\label{fig:cmprelerr} The probability distribution for the relative error $\vep^f_r =
|f^t_{12}-f^m_{12}|/|f^t_{12}|$. }
\efig
A prediction for the force factor $f_{12}$ can be made by combining the two ML models obtained in Section
\ref{sect:ml_model}. In this section the predictions $f_{12}^m$ made this way are compared with the
true predictions $f_{12}^t$ found from solving Eqs. \ref{eq:km_mettin} and \ref{eq:bfctr}. 
$1024$ samples for $(D, R_{E1}, R_{E2}, p_a, \om)^T$ are randomly chosen, with each component
falling in the range covered by the dataset used to train the ML models. Note that the samples are not
necessarily in the dataset. Excluding the samples where $R_{E1}<R_{E2}$, $640$ samples are used for
this test, and $640$ pairs of values $(f^t_{12}, f^m_{12})$ are obtained.  

Excellent correlation is found between $f^t_{12}$ and $f^m_{12}$, with the correlation coefficient 
being $0.98$. 
The scatter plot for the data is shown in Fig. \ref{fig:cmpscat}. 
The figure confirms the good correlation while, in the meantime, shows that 
the difference tends to increase when the magnitude of the force increases. 

The histogram for the relative error $\vep^f_r \equiv |f^t_{12}-f^m_{12}|/|f^t_{12}|$ is shown in
Fig. \ref{fig:cmprelerr}. The error distribution has a broader spread than those in Fig.
\ref{fig:NNf12err}, i.e., those for the training data and the testing data. This behaviour is not
unexpected as the ML models are being applied to a new dataset here. Nevertheless, more than $45\%$
samples have less than $20\%$ errors, and for more than $75\%$ samples the error is less
than $40\%$. Therefore, the results are still quite satisfactory. Obviously, the performance of the
ML models can be improved if the training data set can be expanded and refined. 

%For this testing dataset, $f^t_{12}$ is found be positive on all samples. Mis-classification is
%indeed observed, leading to prediction of negative $f^m_{12}$ for a handful samples.   

Finally, the ML models are extremely efficient compared with direct numerical integration. It takes
about one hour to obtain the $640$ values for $f^t_{12}$, whereas it takes less than one second
for the ML models to find corresponding $f^m_{12}$.  

\section{Conclusions \label{sect:conclusions}}

Machine learning models for 
the secondary Bjerknes force as a function for several parameters have been developed in a two bubble system. Because
the force varies drastically with the parameters,
the magnitude and the sign of the force have to be
modelled separately, which results in a composite model consisting of a feed-forward neural network
for (the logarithm of) the former and a support-vector machine for the latter. 

Numerical tests demonstrate the feasibility of using machine
learning to tackle this problem. Practical methods for choosing the suitable architecture and
hyperparameters for the models are proposed.
Accurate machine models are obtained, which are shown to be very efficient compared with direct numerical
integration of the bubble evolution equations. 

The results demonstrate that machine learning is a viable method in modelling 
the interactions between bubbles. 
The models developed here have the potential to enhance 
the future simulations of bubble clusters. Obviously, the model can be further refined and expanded, by, e.g.,
using a larger dataset 
covering a wider range of parameters. 
Machine learning clearly is equally applicable when 
a more sophisticate physical model is used to describe bubble oscillations, although it is not obvious
that the currently chosen architectures are still sufficient when the dataset grows
larger. These interesting
topics will be explored in our future investigations. 

\section{Acknowledgement}
The authors gratefully acknowledge the support provided by  
the Guangzhou Science (Technology) Research Project (Project No.
201704030010) and the special fund project of science and technology innovation strategy of Guangdong Province.

\bibliographystyle{elsarticle-harv}
\bibliography{./turbref}

\end{document}